\documentclass[11pt]{article}

\usepackage{epsfig}
\usepackage{subfig}
\usepackage{graphicx}
\usepackage{epstopdf}
\usepackage{amsfonts}
\usepackage{amssymb}
\usepackage{amsthm}
\usepackage{amsmath}
\usepackage{multirow}
\usepackage{color}
\usepackage{cite}
 \usepackage{makecell}
 \usepackage{slashed}
\def\beq{\begin{equation}}
\def\eeq{\end{equation}}
\def\bea{\begin{eqnarray}}
\def\eea{\end{eqnarray}}
\newcommand{\beqs}{\begin{subequations}}
\newcommand{\eeqs}{\end{subequations}}

\newcommand{\cref}[1]{Ref.~\cite{#1}}

\newcommand{\hh}{{\ensuremath{I{\kern-2.6pt h}}}}
\newcommand{\bhh}{{\ensuremath{\bar{I{\kern-2.6pt h}}}}}

\usepackage[colorlinks=true,allcolors=blue]{hyperref}
\usepackage{mathtools}

\usepackage[thinc]{esdiff}

\begin{document}

\begin{titlepage}
	

\begin{center}
{\Large {\bf Superheavy Metastable Strings in SO(10)}
}
\\[12mm]
Rinku Maji,$^{1}$
Qaisar Shafi$^{2}$~
\end{center}
\vspace*{0.50cm}
	\centerline{$^{1}$ \it
		Cosmology, Gravity and Astroparticle Physics Group, Center for Theoretical Physics of the Universe,}
		\centerline{\it  Institute for Basic Science, Daejeon 34126, Republic of Korea}
	\vspace*{0.2cm}
	\centerline{$^{2}$ \it
		Bartol Research Institute, Department of Physics and 
		Astronomy,}
	\centerline{\it
		 University of Delaware, Newark, DE 19716, USA}
	\vspace*{1.20cm}
\begin{abstract}
The spontaneous breaking of $SO(10)$ grand unified symmetry to $SU(3)_c \times SU(2)_L \times U(1)_Y \times U(1)_\chi$ yields the GUT monopole as well as a comparably heavy $U(1)_\chi$ monopole which also carries $U(1)_Y$ flux. A metastable string scenario in this case requires that the $U(1)_\chi$ symmetry is necessarily broken close to the GUT scale, thus resulting in a dimensionless string tension $G \mu \sim 10^{-6}$. We show that the $\chi$ monopole does not carry any unconfined flux following the electroweak symmetry breaking. Coupled with $G \mu \sim 10^{-6}$, this metastable string network appears to provide a good fit to the recent Pulsar Timing Array data on the stochastic gravitational background.
Gauge coupling unification, especially in the presence of low scale supersymmetry, determines the GUT scale and, in combination with constraints from proton decay experiments, one is able to constrain some of the key parameters in this setup.
The breaking of $SO(10)$ via $SU(5) \times U(1)_\chi$ also yields superheavy metastable strings with no unconfined flux associated with the monopoles. Finally, we consider $SO(10)$ breaking via $SU(4)_c \times SU(2)_L \times U(1)_R$, $SU(3)_c \times SU(2)_L \times SU(2)_R \times U(1)_{B-L}$ and flipped $SU(5)$ that yield metastable strings where the associated monopoles carry unconfined flux after the electroweak breaking.
\end{abstract}

\end{titlepage}
\section{Introduction}
The gravitational radiation emitted by superheavy metastable \cite{Buchmuller:2021mbb, NANOGrav:2023hvm,Buchmuller:2023aus} or quasistable \cite{Lazarides:2022jgr, Lazarides:2023ksx} cosmic strings, with a dimensionless string tension $G \mu \sim 10^{-6}$, provides a plausible explanation of the stochastic gravitational background reported by the various Pulsar Timing Array (PTA) experiments \cite{NANOGrav:2023gor,Antoniadis:2023ott,Reardon:2023gzh,Xu:2023wog}. The presence of the composite structures `monopoles connected by strings' is essential to form such string networks. In the metastable scenario the primordial monopoles are inflated away, and the strings which confine the flux of the monopoles, partially or completely, are formed in the subsequent symmetry breaking. Quantum mechanical tunneling of monopole-antimonopole pairs on these strings eventually causes the strings to disappear. In the quasistable scenario on the other hand, the quantum mechanical tunneling of monopoles is suppressed, and so it’s the horizon reentry, following inflation, of the primordial monopoles that eventually causes the strings to form dumbbells and finally decay. The stochastic gravitational wave spectra predicted in the two scenarios is compatible with the current observations.

It is well known \cite{Kibble:1982ae} that topologically stable strings appear in $SO(10)$ \cite{Georgi:1974my,Fritzsch:1974nn}, more precisely Spin(10), if this symmetry is broken to the Standard Model (SM) and subsequently to $SU(3)_c \times U(1)_{\rm em}$, by employing Higgs fields in the tensor representations.

Metastable cosmic strings in $SO(10)$ have attracted a fair amount of recent attention, with the discussion mostly focused on the breaking of $SO(10)$ via its left-right symmetric subgroups \cite{Antusch:2023zjk,Antusch:2024nqg, Maji:2024tzg} (For other related studies see Refs.~\cite{Lazarides:2023rqf,Maji:2023fhv,Ahmed:2023rky, Afzal:2023cyp,Fu:2023mdu, Ahmed:2023pjl,King:2023wkm,Afzal:2023kqs,Maji:2024cwv,Pallis:2024joc,Maji:2024pll,Chitose:2024pmz,Pallis:2024mip,Datta:2024bqp, Ahmad:2025dds,Antusch:2025xrs}). Somewhat surprisingly perhaps, the appearance of metastable strings from the breaking of $SO(10)$ to the SM via $SU(5) \times U(1)_\chi$ or via $SU(3)_c \times SU(2)_L \times U(1)_Y \times U(1)_\chi$, has drawn less attention. In Ref.~\cite{Lazarides:2023iim} it was shown that the breaking of $SO(10)$ via $SU(5) \times U(1)_\chi$ to the SM produces a dumbbell configuration with an $SO(10)$ monopole-antimonopole pair at the two ends. Assuming that the $SO(10)$ breaking scale is close to $M_{\rm GUT}$, this breaking chain certainly provides a viable metastable string scenario if the $U(1)_\chi$ symmetry is also broken close to $M_{\rm GUT}$.
In this paper we propose an alternative scheme where $SO(10)$ is directly broken to $SU(3)_c \times SU(2)_L \times U(1)_Y \times U(1)_\chi$ which yields two superheavy monopoles, namely the well-known GUT monopole, as well a $U(1)_\chi$ monopole of comparable mass that also carries some $U(1)_Y$ flux. Implementation of the metastable string scenario in this case requires that the $U(1)_\chi$ symmetry is necessarily broken close to the GUT scale such that the dimensionless string tension $G\mu$ is predicted to be on the order of $10^{-6}$. We also show that following the electroweak symmetry breaking, the entire $\chi$ monopole flux is squeezed into a tube with no unconfined (electromagnetic) flux. It has been shown in the literature \cite{NANOGrav:2023hvm} that superheavy metastable cosmic strings without unconfined flux provide a nice fit to the PTA data.
It is important to note that gauge coupling unification in this breaking scheme can be elegantly realized without relying on an intermediate gauge symmetry breaking scale. In a supersymmetric setting this is a well-known result \cite{Ellis:1990wk,Amaldi:1991cn,Langacker:1991an,Amaldi:1991zx,Chakrabortty:2017mgi}, and in a non-supersymmetric framework, successful unification of the SM gauge couplings can be implemented by introducing a suitable set of vectorlike quarks in the TeV scale \cite{Frampton:1983sh}.
The breaking of $SO(10)$ gauge symmetry via $SU(5) \times U(1)_\chi$, also yield a similar metastable string network, with the monopole flux confined after the electroweak symmetry breaking. Finally, we present two examples of $SO(10)$ breaking via subgroups of $SU(4)_c\times SU(2)_L\times SU(2)_R$ \cite{Pati:1974yy}, and one based on flipped $SU(5)$ that yield a metastable network in which the accompanying monopoles carry unconfined flux after the electroweak breaking. Future experiments hopefully can determine if any of these scenarios is realized in nature.
\section{Metastable string with no unconfined flux}
\label{sec:2}
In $SO(10)$ grand unification the 15 chiral fermions of each family reside, together with a right handed neutrino, in the 16-dimensional spinor representation. Under the decomposition $SO(10) \to SU(5) \times U(1)_\chi \to SU(3)_c \times SU(2)_L \times U(1)_Y$, we obtain
\begin{align}
& 16 = 10(-1) + \bar{5}(3) + 1(-5) ,  \quad
10 =  (3,2)(1) + (\bar{3}, 1)(-4) + (1,1)(6) \nonumber \\ 
& \overline{5} = (\overline{3},1)(2) + (1,2)(-3).
\end{align}
Following Ref.~\cite{Slansky:1981yr}, the $U(1)$ generators are normalized such that they have the minimal integer charges compatible with a period of $2\pi$. Note that according to this normalization the electric charge generator is
\begin{align}
\label{eq:Q}
Q=T^3_L/2+Y/6 .
\end{align}

To implement the metastable cosmic string scenario we consider the following $SO(10)$ symmetry breaking:
\begin{align}
\label{eq:breaking-chain}
SO(10)\to H = SU(3)_c \times SU(2)_L \times U(1)_Y \times U(1)_\chi & \to SU(3)_c \times SU(2)_L \times U(1)_Y \nonumber \\ &\to SU(3)_c \times U(1)_{\rm em}.
\end{align}
The breaking of $SO(10)$ to $H$ which contains two $U(1)$ factors means that two varieties of superheavy monopoles are present at this stage. To identify the monopoles with the minimal charges we should study the first homotopy group of $H$. The task is made easier by recognizing that the topologically stable superheavy GUT monopole arises from  $SU(5)$ which contains the electric charge generator $Q$. The minimal charge GUT monopole, it turns out, carries a single quantum ($2\pi/e$) of Dirac magnetic charge as well as some color magnetic charge which is screened. This can be seen realizing that $U(1)_Y$ intersects both $SU(3)_c$ and $SU(2)_L$ in their respective centers. Thus, we can create a non-trivial loop in the SM group space by doing a $2\pi/6$ rotation along $U(1)_Y$, followed by a $2\pi/3$ rotation along the color hypercharge generator $T_c^8={\rm diag}(1,1,-2)$, and a rotation by $2\pi$ along $T_L^3/2$. Since $Q = T_L^3/2 + Y/6$, we have  identified a magnetic monopole associated with a $2\pi$ rotation along $Q$, which corresponds to a Dirac magnetic charge of $2\pi/e$. The monopole also carries some color magnetic flux which is screened. 

\begin{figure}[htbp]
\begin{center}
\includegraphics[scale=0.75]{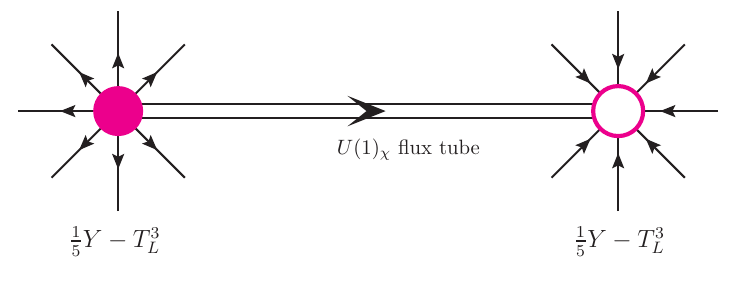}
\caption{$\chi$ monopole-antimonopole pair connected by $U(1)_\chi$ flux tube. Electroweak symmetry is not yet broken.}
\label{fig:3211-1}
\end{center}
\end{figure}
\begin{figure}[htbp]
\begin{center}
\includegraphics[scale=0.75]{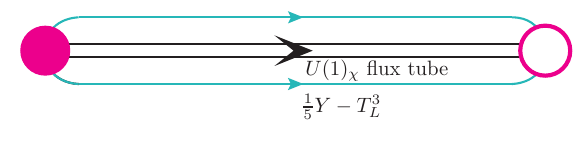}
\caption{Following electroweak breaking the flux $Y/5 - T_L^3$ in Figure \ref{fig:3211-1} is also squeezed inside a tube.}
\label{fig:3211-2}
\end{center}
\end{figure}
Next, we are interested in the second monopole arising from the presence of $U(1)_\chi$, referred to here as the $\chi$ monopole. We call it a $\chi$ monopole because all matter fields in the 16-plet interact with it, including the right handed neutrino which is a SM as well as an $SU(5)$ singlet. The minimal charge for this monopole is found by realizing \cite{Lazarides:2023iim} that as topological spaces, $U(1)_\chi$ and $U(1)_Y$ intersect in $Z_5$, the center of $SU(5)$. Thus, we can perform a rotation by $2 \pi/5$ along $U(1)_\chi$, and then return to the identity element with a $2\pi/5$ rotation along $U(1)_Y$. In other words, the minimal $\chi$ monopole simultaneously carries both $U(1)_\chi$ and $U(1)_Y$ flux. Since the unbroken symmetry at this stage is $H$, we can include a $-2 \pi$ rotation in the simply connected space $SU(2)_L$ with the generator $T_L^3 ={\rm diag}(1,-1)$.
 The overall rotation corresponding to this minimal monopole is then given by
\begin{align}
\label{eq:chi-mon}
\exp\left\lbrace\frac{2\pi i}{5}(\chi + Y - 5 T_L^3)\right\rbrace
\end{align}
 The breaking of $U(1)_\chi$ is done with a $\nu^c$-type Higgs field in the 16-dimensional representation of $SO(10)$. Following the breaking of $U(1)_\chi$ close to $M_{\rm GUT}$, the $\chi$ magnetic flux is squeezed into a tube but the electroweak magnetic flux is still in a Coulomb phase. However, the spontaneous breaking of the electroweak symmetry also squeezes this Coulomb flux because it corresponds to the broken generator orthogonal to the electric charge $Q$. Since $Q = T_L^3 / 2 + Y/6$, the orthogonal broken generator is 
\begin{align}
\label{eq:B}
\mathcal{B} = {T_L^3}/{2} -{Y}/{10} .
\end{align}
This argument shows that the dumbbell formed by the $\chi$ monopole-antimonopole pair and joined by a flux tube does not carry any Coulomb flux (see Figure \ref{fig:3211-2}). We can verify from Eq.~\eqref{eq:chi-mon} that the electroweak flux of the $\chi$ monopole is confined inside a tube by taking, for instance, the neutrino and electron with respective $\mathcal{B}$ charges of $8/10$ and $-2/10$, around it.  Here, $\chi + Y-5T^3_L=\chi - 10\mathcal{B}$ is 5 or $-5$ for all fermions in the SM, so that we return to identity around the string, according to Eq.~\eqref{eq:chi-mon}.
Thus, we have realized a superheavy metastable string scenario if $SO(10)$ symmetry breaking proceeds via the subgroup $H=SU(3)_c \times SU(2)_L \times U(1)_Y \times U(1)_\chi$. Finally, since $U(1)_\chi$ is spontaneously broken by the vacuum expectation value of a SM singlet scalar field, we have made the plausible assumption that the fundamental (GUT) monopole does not carry any $\chi$ flux.

 Superheavy metastable strings also appear if the $SO(10)$ breaking to the SM proceeds as follows:
\begin{align}
\label{eq:breaking-2}
SO(10) \to SU(5) \times U(1)_\chi \to SU(3)_c \times SU(2)_L \times U(1)_Y.
\end{align}
In contrast to our previous discussion the $SO(10)$ $\chi$ monopole mass is now linked to a scale ($M_{10}$) that is larger than the usual GUT symmetry breaking scale. To implement the metastable string scenario, the mass scale $M_{10}$ cannot be much larger than the second ($SU(5)$) symmetry breaking scale $M_{\rm GUT}$, which also is the scale for $U(1)_\chi$ breaking that yields the metastable string scenario. Based on recent progress in the construction of realistic hybrid inflation models \cite{Lazarides:2023rqf, Moursy:2024hll,Maji:2024cwv},  this scenario, with $SU(5)$ and $U(1)_\chi$ broken at comparable scales, offers the possibility of realizing an observable number density of primordial GUT ($SU(5)$) monopoles as well as gravitational waves from the superheavy metastable string network.
The primordial $\chi$ monopole, of course, is inflated away.
Proton decay offers another way to distinguish between the two symmetry breaking schemes in Eq.~\eqref{eq:breaking-chain} and Eq.~\eqref{eq:breaking-2}.
Clearly with $M_{10}> M_{\rm GUT}$, proton decay processes mediated by the gauge bosons outside of $SU(5)$ will be suppressed by an order of magnitude or two, depending on how much larger $M_{10}$ is compared to $M_{\rm GUT}$. 

\section{Metastable strings with unconfined flux}
\label{sec:MSS-uncon-flux}
In this section we describe some $SO(10)$ breaking chains that provide metastable strings where the confined monopoles and antimonopoles also carry unconfined electromagnetic magnetic flux.
Consider the breaking
\begin{align}
\label{eq:421}
SO(10) &\to SU(4)_c \times SU(2)_L \times U(1)_R \nonumber \\
&\to SU(3)_c \times U(1)_{B-L} \times SU(2)_L \times U(1)_R \to SU(3)_c \times SU(2)_L \times U(1)_Y .
\end{align}
The first breaking produces a monopole which turns out to be the topologically stable GUT monopole.
\begin{figure}[htbp]
\begin{center}
\includegraphics[scale=0.75]{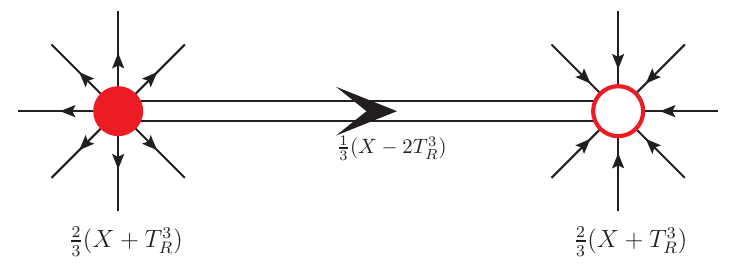}
\caption{Monopole-antimonopole pair connected by a string with unconfined flux during the symmetry breaking given in Eq.~\eqref{eq:421} (before electroweak breaking).}
\label{fig:421-1}
\end{center}
\end{figure}
\begin{figure}[htbp]
\begin{center}
\includegraphics[scale=0.75]{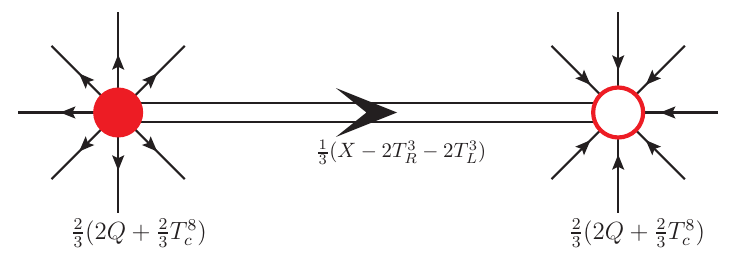}
\caption{Monopole-antimonopole pair connected by a string plus unconfined Coulomb flux and color magnetic flux after the electroweak symmetry breaking for the chain given in Eq.~\eqref{eq:421}.}
\label{fig:421-2}
\end{center}
\end{figure}
The second breaking produces an $SU(4)_c$ monopole, referred to as a ‘red’ monopole in Refs.~\cite{Lazarides:2019xai,Lazarides:2023iim}.  This monopole carries one unit of $X$ flux, where $X$ is the $SU(4)_c$ generator $(B-L) + \frac{2}{3}T_c^8$. To check this, note that a $2 \pi$ rotation along $(B-L)$ followed by a $2 \pi$ rotation along $\frac{2}{3} T_c^8$ yields a non-trivial closed loop in $SU(3)_c \times U(1)_{B-L}$. We can add to this a $2\pi$ rotation along $T_L^3$ since $SU(2)_L$ is unbroken.
The third step in Eq.~\eqref{eq:421} involves the breaking of $U(1)_{B-L} \times U(1)_R$ to $U(1)_Y$. The broken generator is $X-2 T_R^3$ \cite{Lazarides:2019xai,Lazarides:2023iim}, and the monopole at this stage with the accompanying string is shown in Figure \ref{fig:421-1}.  Note that the breaking of $X -2 T_R^3$ is achieved with the $\nu^c$ type field in the Higgs 16-plet.
 
 In Figure \ref{fig:421-2} we display the configuration following the electroweak symmetry breaking. Clearly, in contrast to the breaking of $SO(10)$ via the breaking chains involving $U(1)_\chi$, the $SU(4)_c$ monopole attached to the string carries some unconfined magnetic flux.

Following Ref.~\cite{Lazarides:2019xai} it can be shown that a similar result holds if the monopole accompanying the metastable string arises from the break of the $SU(2)_R$ symmetry. The monopole-string system in this case too is accompanied by some unconfined magnetic flux. This is realized, for instance, in the following symmetry breaking scenario:
\begin{align}
\label{eq:3221}
SO(10) &\to SU(3)_c \times SU(2)_L \times SU(2)_R\times U(1)_{B-L} \nonumber \\
&\to SU(3)_c \times SU(2)_L \times U(1)_{B-L} \times U(1)_R \to SU(3)_c \times SU(2)_L \times U(1)_Y .
\end{align}
\begin{figure}[htbp]
\begin{center}
\includegraphics[scale=0.75]{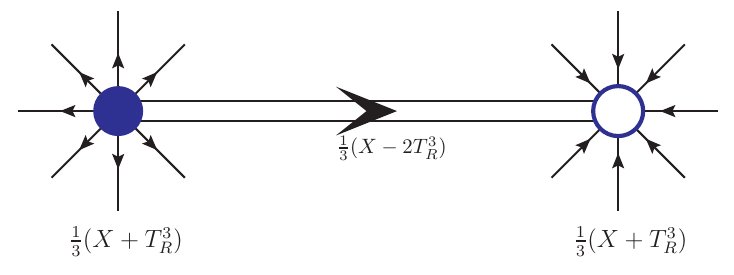}
\caption{Monopole-antimonopole pair connected by a string with unconfined fluxes during the symmetry breaking given in Eq.~\eqref{eq:3221} (before electroweak breaking).}
\label{fig:3221-1}
\end{center}
\end{figure}
\begin{figure}[htbp]
\begin{center}
\includegraphics[scale=0.75]{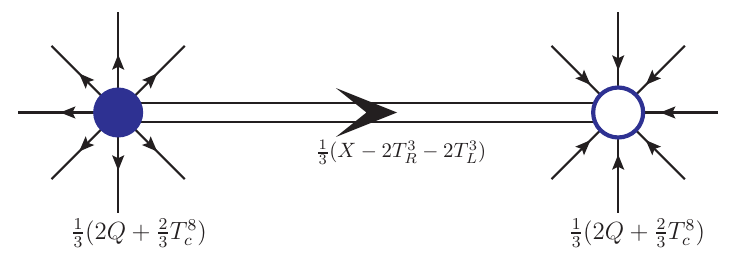}
\caption{Monopole-antimonopole pair connected by a string with unconfined Coulomb flux and color magnetic flux after the electroweak symmetry breaking for the chain given in Eq.~\eqref{eq:3221}.}
\label{fig:3221-2}
\end{center}
\end{figure}
As shown in Ref.~\cite{Lazarides:2019xai} the first breaking produces the topologically stable GUT monopole. The $SU(2)_R$ breaking to $U(1)_R$ in the second step produces the (blue) monopole, whose flux is partially squeezed into a tube in the third step of symmetry breaking, as shown in Figure \ref{fig:3221-1}. The final monopole-string system with the unconfined magnetic flux is shown in Figure \ref{fig:3221-2}.

This discussion can be extended to other related scenarios such as $SO(10)$ breaking via the gauge symmetry $SU(4)_c \times SU(2)_L \times SU(2)_R$ \cite{Pati:1974yy}. This offers the possibility of creating two sets of cosmic metastable strings associated with the $SU(4)_c$ (red) and $SU(2)_R$ (blue) monopoles. Furthermore, the merger of the red and blue monopoles yields topologically stable monopoles that carry two units
($4 \pi/e$) of Dirac magnetic charge as well as color magnetic charge \cite{Lazarides:2019xai,Lazarides:2023iim,Lazarides:2024niy,Kephart:2025tik} which presumably is screened.

 For completeness, we briefly describe here the appearance of superheavy metastable strings from the breaking of $SO(10)$ via the symmetry breaking chain (via flipped $SU(5)$):
\begin{align}
\label{eq:fsu5-chain}
SO(10) \to SU(5)\times U(1)_X & \to SU(3)_c \times SU(2)_L \times U(1)_Z \times U(1)_X \nonumber \\ & \to SU(3)_c \times SU(2)_L \times U(1)_Y.
\end{align}
The first breaking in Eq.~\eqref{eq:fsu5-chain} produces a monopole that carries $U(1)_X$ and $SU(5)$ magnetic flux. It evolves into the topologically stable GUT monopole if we recall that the electric charge generator $Q$ receives contribution from the generator of $U(1)_X$.
The confined monopole associated with the metastable string arises from the breaking of $SU(5)$ to $SU(3)_c \times SU(2)_L \times U(1)_Z$ and it carries magnetic flux associated with these three groups. In the third breaking the unbroken ($Y$) and broken ($\tilde{Y}$) generators are respectively given by
\begin{align}
\label{eq:fsu5-gentrs}
Y = -(Z + 6 X)/5, \quad \tilde{Y} = (-4 Z + X)/5.
\end{align}
The breaking of $U(1)_Z \times U(1)_X$ to $U(1)_Y$ is implemented with a $\nu^c$ type Higgs field in the 16-plet of SO(10). The symmetry breaking $U(1)_Z \times U(1)_X \to U(1)_Y$ produces a flux tube corresponding to the broken generator $\tilde{Y}$ in Eq.~\eqref{eq:fsu5-gentrs}. 

\begin{figure}[htbp]
\begin{center}
\includegraphics[scale=0.75]{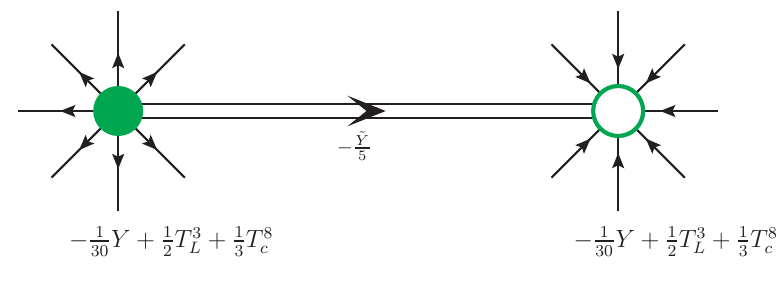}
\caption{Dumbbell configuration following the breaking of $U(1)_Z \times U(1)_X$ to $U(1)_Y$, which confines the $Z$ monopole-antimonopole pair.}
\label{fig:fsu5-1}
\end{center}
\end{figure}
\begin{figure}[htbp]
\begin{center}
\includegraphics[scale=0.75]{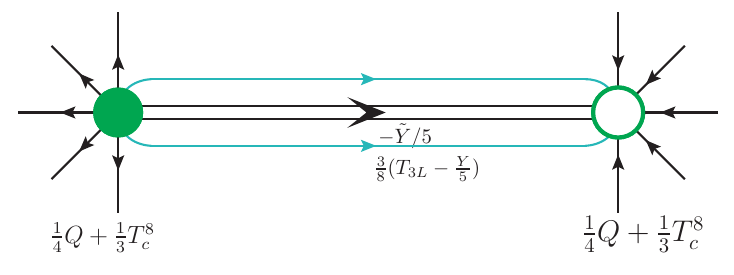}
\caption{Dumbbell configuration electroweak symmetry breaking. The electroweak flux is only partially confined in the breaking of $SO(10)$ via flipped $SU(5)$.}
\label{fig:fsu5-2}
\end{center}
\end{figure}
In Figures \ref{fig:fsu5-1} and \ref{fig:fsu5-2} we display the dumbbell configuration for this case. The confined monopole flux in this case corresponds to a non-trivial closed loop in $SU(3)_c \times SU(2)_L \times U(1)_Z$, which is obtained by performing a $2 \pi/6$ rotation in $U(1)_Z$, followed by a $\pi$ rotation along $T_L^3$ and a $2 \pi$ rotation along $\frac{1}{3} T_c^8$. The monopole-antimonopole configuration following the breaking of the symmetry $U(1)_Z \times U(1)_ X \to U(1)_Y$ is shown in Figure \ref{fig:fsu5-1}. After electroweak breaking, the monopole-antimonopole pair ends up carrying some Coulomb flux since $-\frac{Y}{30} =
\frac{1}{4} Q +\frac{3}{4} \mathcal{B}$.
The system also carries some color magnetic charge which presumably is  screened.

To summarize, we have identified in this section some prominent $SO(10)$ symmetry breaking chains that yield superheavy metastable cosmic strings with unconfined magnetic flux.

\section{Proton decay, gravitational waves, and PTA data}
\label{sec:3}
The superheavy gauge bosons with the SM quantum numbers $(3,2,-5)$ and $(3,2,1)$ mediate proton decay. The relevant dimension six operators after integrating out these fields are given \cite{FileviezPerez:2004hn, Nath:2006ut} in the physical basis as
        \begin{align}
            \label{operator_physical_basis}
             \mathcal{W_C} \left( \epsilon^{ijk} \overline{u^c_i}\gamma^\mu u_j \overline{e^c} \gamma_\mu d_{k} + \epsilon^{ijk}
            \overline{u^c_i}\gamma^\mu u_j \overline{d^c_{k}} \gamma_\mu e\right),
        \end{align}
\begin{figure}[h!]
\begin{center}
\includegraphics[scale=0.9]{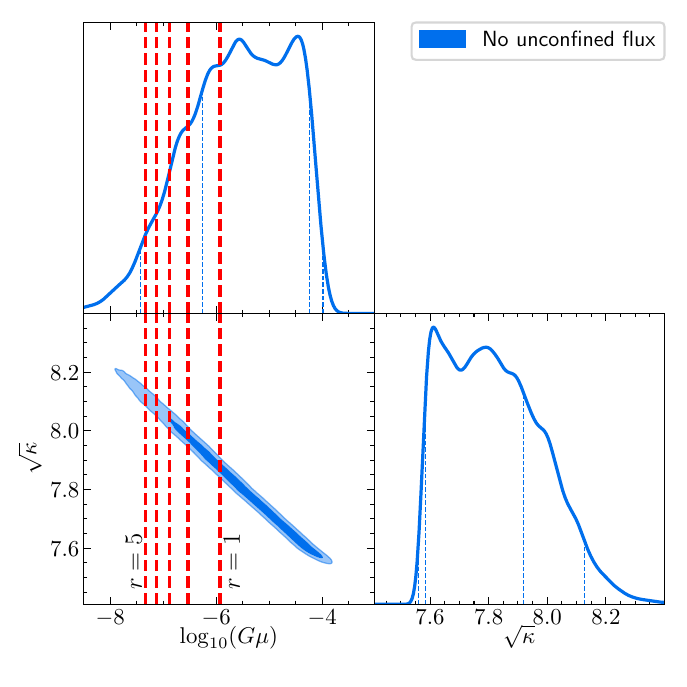}
\end{center}
\caption{Posterior distribution of $G\mu$ and $\sqrt{\kappa}$ for metastable strings with no unconfined flux based on the NANOGrav 15 year data \cite{NANOGrav:2023hvm,nanograv_2023}. The dark and light blue regions in the off-diagonal plot depict $1\sigma$ and $2\sigma$ Bayesian credible regions. The diagonal plots show the marginalized 1D distributions and blue dashed vertical lines indicate 68\% and 95\% credible intervals. The red dashed lines depict the lower bound, given by Eq.~\eqref{eq:gmu-taup}, on the string tension $G\mu$ from the proton lifetime bound of Super-Kamiokande experiment for $r$ values from 1 to 5. The higher $G\mu$ values on the right of the dashed red line is ruled out from the bound on the proton lifetime for a given $r$ for the breaking chain in Eq.~\eqref{eq:breaking-chain}.}
\label{fig:Bayes-MSS}
\end{figure}
        where 
        \begin{align}
        \label{eq:WC}
        \mathcal{W_C}=\frac{(1+|V_{ud}|^2)}{2 v_{\rm GUT}^2}
        \end{align}
        is the Wilson coefficient and $|V_{ud}| = 0.9742$ is the CKM matrix element \cite{ParticleDataGroup:2022pth}. Here $v_{\rm GUT}$ denotes the vev associated with the breaking $SO(10)\to SU(3)_c\times SU(2)_L\times U(1)_Y\times U(1)_\chi$ which creates the $\chi$ monopoles. The partial decay lifetime for the channel $p\to \pi^0 e^+$
        is given by \cite{FileviezPerez:2004hn}
        \begin{align}\label{eq:taup-1}
            {\tau_p} =\left[\frac{m_p}{32\pi}\left(1-\frac{m_{\pi^0}^2}{m_p^2}\right)^2 A_L^2\mathcal{W_C}^2 
         \left( A_{SL}^2 |\langle \pi^0 \rvert (ud)_L u_L\lvert p \rangle |^2 +
            A_{SR}^2 |\langle \pi^0 \rvert (ud)_R u_L\lvert p \rangle |^2
         \right)\right]^{-1},
        \end{align}
        where $m_p$ and $m_{\pi^0}$ are the proton and neutral pion masses respectively.
    The quantities $A_{SR(L)}$ \cite{Buras:1977yy,Abbott:1980zj} and $A_L\simeq 1.33$ \cite{Nihei:1994tx}, namely the long and short range enhancement factors, arise from the renormalization group evolution of the proton decay operators from the GUT scale to the electroweak scale and then to the QCD scale ($\sim 1$ GeV). We use the form-factors from the lattice QCD computation \cite{Aoki:2017puj}:
        \begin{align}
        \langle \pi^0 \rvert (ud)_R u_L\lvert p \rangle  = -0.131, \ \ \langle \pi^0 \rvert (ud)_L u_L\lvert p \rangle = 0.134 \ .
        \end{align}        
        
 The string tension is $\mu \simeq \pi v_s^2$, where $v_s$ denotes the vev associated with the breaking of $U(1)_\chi$. We can express the Wilson coefficient in Eq.~\eqref{eq:WC} as 
 \begin{align}
 \mathcal{W_C} = \frac{(1+|V_{ud}|^2)}{16 m_{\rm Pl}^2 r^2 G\mu}
\end{align} where the ratio $r=v_{\rm GUT}/v_s$ and $m_{\rm Pl}^2 = 1/(8\pi G)$. The partial lifetime in Eq.~\eqref{eq:taup-1} can be written as
\begin{align}\label{eq:taup-2}
\tau_p = 1.72\times 10^{45}(G\mu)^2r^4~\mathrm{yrs.}
\end{align}
 The constraint $\tau_p>2.4\times 10^{34}$ yrs from the Super-Kamiokande experiment \cite{Super-Kamiokande:2020wjk} gives 
\begin{align}\label{eq:gmu-taup}
G\mu\gtrsim 1.2\times 10^{-6}/r^2
\end{align} 
The decay width per unit length of the string is given by
\begin{align}
\Gamma_s = \frac{\mu}{2\pi}\exp(-\pi\kappa),
\end{align}
where $\kappa=m_M^2/\mu$ is the metastability factor and $m_M$ is the monopole mass. In Figure \ref{fig:Bayes-MSS}, we have shown the posterior  distribution of $G\mu$ and $\sqrt{\kappa}$ for metastable strings with no unconfined flux with the NANOGrav 15 year data \cite{NANOGrav:2023hvm,nanograv_2023}. We have used the the Markov Chain Monte Carlo chain available in Zenodo repository \cite{nanograv_2023}. The proton lifetime constraint from the Super-Kamiokande experiment provides a lower bound on $G\mu$ which is shown with the red dashed lines for $r$ values $1-5$. 

\section{Conclusions}
\label{sec:conc}
We have shown that the breaking of $SO(10)$ to the Standard Model via the gauge symmetry $SU(3)_c \times SU(2)_L \times U(1)_Y \times U(1)_\chi$ offers some unique features which can be experimentally tested in a number of ways. First, it yields metastable cosmic strings with a dimensionless string tension $G \mu$  predicted to be of order $10^{-6}$. Second, the flux tube associated with the metastable string does not carry any unconfined flux which, combined with the predicted dimensionless string tension, provides a compelling fit to the PTA data. Third, gauge coupling is nicely implemented in the presence of low scale supersymmetry without any additional assumptions. Last but not least, observable proton decay remains a key prediction of this model.
We also briefly discuss the breaking of $SO(10)$ via $SU(5) \times U(1)_\chi$. In  $SU(5) \times U(1)_\chi$, metastable strings with slightly higher $G \mu$ values than $10^{-6}$ are possible, and the monopole does not carry unconfined flux.  Finally, we identify $SO(10)$ breaking schemes via the gauge symmetry $SU(4)_c\times SU(2)_L\times SU(2)_R$ and its subgroups, and also via flipped $SU(5)$, that yield metastable strings with the monopoles carrying unconfined flux.
\section{Acknowledgment}
R.M. is supported by the Institute for Basic Science under the project code: IBS-R018-D3.
\appendix

\bibliographystyle{JHEP}
\bibliography{Cparity_cleaned}

\end{document}